\documentclass[reprint]{revtex4-2}

\usepackage{graphicx}
\usepackage{dcolumn}
\usepackage{bm}
\usepackage{hyperref}
\usepackage{physics}
\usepackage{xcolor}
\newcommand{\etal}{\textit{et al.}}
\begin{document}
\title{Entanglement in first excited states of some many-body quantum spin systems:\\indication of quantum phase transition in finite size systems}
\author{George Biswas}
\author{Anindya Biswas}
\affiliation{Department of Physics, National Institute of Technology Sikkim\\ Ravangla, South Sikkim 737139, India.}
\date{\today}
\begin{abstract}
We compute concurrence, a measure of bipartite entanglement, of the first excited state of the $1$-D Heisenberg frustrated $J_1$-$J_2$ spin-chain and observe a sudden change in the entanglement of the eigen state near the coupling strength $\alpha=J_2/J_1\approx0.241$, where a quantum phase transition from spin-fluid phase to dimer phase has been previously reported. We numerically observe this phenomena for spin-chain with $8$ sites to $16$ sites, and the value of $\alpha$ at which the change in entanglement is observed, asymptotically tends to a value $\alpha_c\approx0.24116$. We have calculated the finite-size scaling exponents for spin chains with even and odd spins. It may be noted that bipartite as well as multipartite entanglement measures applied on the ground state of the system, fail to detect any quantum phase transition from the gapless to the gapped phase in the $1$-D Heisenberg frustrated $J_1$-$J_2$ spin-chain. Furthermore, we measure bipartite entanglement of first excited states for other spin models like $2$-D Heisenberg $J_1$-$J_2$ model and Shastry-Sutherland model and find similar indications of quantum phase transitions.
\end{abstract}
\maketitle
\section{INTRODUCTION}
\label{sec1}
Advances in technology in the field of low temperature experiments have made it possible to engineer some quantum many-body Hamiltonians using ultracold atoms and ions~\cite{PhysRevLett.120.243201}. Such quantum spin systems may be important as substrates for quantum computation. Quantum entanglement is a resource for quantum computational tasks. Therefore, it is important to study and understand entanglement in such systems. Bipartite and multipartite entanglement~\cite{RevModPhys.81.865,PhysRevA.81.012308,PhysRevLett.106.190502,Schwaiger2017RelationsBB,PhysRevA.65.032314,RevModPhys.80.517} in ground states of quantum spin systems have been studied and critical quantum phenomena~\cite{PhysRevA.90.032301,RevModPhys.82.277,Calabrese_2009,Latorre:2004:GSE:2011572.2011576,PhysRevLett.94.227002,PhysRevA.66.032110,Latorre_2009} have been detected. However, entanglement of low lying excited states of quantum spin systems have not been exhaustively studied~\cite{requardt2006entanglement,Alba_2009,PhysRevA.80.052104}. In this paper, we compute a nearest neighbor bipartite entanglement measure namely concurrence~\cite{PhysRevLett.78.5022} of qubits in first excited states of some non-integrable quantum spin systems.
 
The systems that we have studied are the one dimensional Heisenberg frustrated $J_1-J_2$ spin chain, the two dimensional Heisenberg $J_1-J_2$ spin system and the Shastry-Sutherland model. The ground states of these systems have been investigated and bipartite and multipartite quantum entanglement have been measured~\cite{PhysRevA.90.032301}. Quantum phase transitions (QPT), a zero temperature phase transition driven by system parameters~\cite{sachdev_2011}, have been detected in some of the cases. However, the quantum phase transition from the spin fluid phase to dimer phase has not been detected using any quantum entanglement measure for the  one dimensional Heisenberg frustrated $J_1-J_2$ spin chain. The quantum phase transition from the gapless phase to the gapped phase in the one dimensional Heisenberg frustrated $J_1-J_2$ spin chain was investigated by Haldane~\cite{PhysRevB.25.4925}, Tonegawa and Harada~\cite{doi:10.1143/JPSJ.56.2153},  Okamoto and Nomura~\cite{OKAMOTO1992433} using exact diagonalization and field theory methods. It was reported that the ground state is in the gapless or gapped phase depending on the value of the coupling strength~$\alpha$. The quantum phase transition point was estimated by investigating the singlet-triplet energy gap of finite size systems~\cite{doi:10.1143/JPSJ.56.2153} followed by extrapolation to infinite system. In Ref.~\cite{OKAMOTO1992433}, the phase transition point was determined by investigating the difference between the singlet-triplet gap and the singlet-singlet gap for finite size systems. In Ref.~\cite{doi:10.1143/JPSJ.56.2153,OKAMOTO1992433}, the singlet-triplet energy gap was defined as
\begin{equation}
 G_{st}(N,\alpha)\equiv E_1^{(0)}(N,\alpha)-E_0^{(0)}(N,\alpha)
 \label{steqn}
\end{equation}
while the singlet-singlet energy gap was defined as
\begin{equation}
 G_{ss}(N,\alpha)\equiv E_0^{(1)}(N,\alpha)-E_0^{(0)}(N,\alpha)
 \label{sseqn}
\end{equation}
where $E_m^{(0)}(N,\alpha)$ and $E_m^{(l)}(N,\alpha)$ are the ground state energy and the $l$th excited state energy in the $S_{total}=m$ subspace, respectively. The first excited states of the systems, considered in this paper, are in general, degenerate. Let $\ket{E_1^i}$ denote the $i$-th degenerate eigenstate corresponding to the eigenenergy $E_1$. Then the density matrix corresponding to the first excited $d$-fold degenerate eigenstate is given by
\begin{equation}
 \rho_1=\frac{1}{d}\sum_{i=1}^{d}\ket{E_1^i}\bra{E_1^i}
 \label{define_state}
\end{equation}
Note that in this paper we do not consider different total spin subspaces explicitly.

We measure the nearest neighbor concurrence of the first excited state $\rho_1$ of the spin chain, and notice a sudden change in the value of concurrence near the quantum phase transition point~\cite{OKAMOTO1992433,PhysRevB.54.R9612}. The computation alongwith the appropriate scaling analysis is done for spin chains consisting of 8 to 16 qubits. The scaling analysis and the corresponding finite size scaling exponents are different for even and odd spin chains. The quantum critical point $\alpha_c\approx0.24116$ is estimated from the scaling analysis of spin chains with even number of qubits. The finite size scaling exponent $\beta=-1.962$. The concurrence versus driving parameter plot Fig.~\ref{fig:conc1d} for spin chains with odd number of qubits shows two discontinuities with both of them converging to the quantum phase transition point in the asymptotic limit. The scaling exponent of the right shifting and left shifting discontinuities are  $\beta_R=-1.92$ and $\beta_L=-2.082$ respectively. The nearest neighbor concurrence of first excited states of the two dimensional Heisenberg  $J_1-J_2$ spin system and the Shastry-Sutherland model for 16 qubits in a ($4\times4$) sites square lattice have also been calculated. We get  indications of quantum phase transition in both the systems.

In Section~\ref{sec2}, we discuss the results for the one dimensional Heisenberg $J_1-J_2$  model in details and highlight the importance of investigating the low lying excited states in quantum spin systems. In Sections~\ref{sec3} and~\ref{sec4} we discuss the results obtained for the two dimensional Heisenberg's $J_1-J_2$ model and Shastry-Sutherland spin model respectively. Finally, we conclude in Section~\ref{sec5}. 
\section{THE one DIMENSIONAL HEISENBERG $J_1-J_2$  SPIN CHAIN}
\label{sec2}
We consider the Heisenberg frustrated one dimensional $J_1-J_2$ model in which the nearest neighbor couplings $J_1$ and the next nearest neighbor couplings $J_2$ are both antiferromagnetic.  The Hamiltonian of the system is given by
\begin{equation}
	H_{1D}=J_1\sum_{i=1}^{N}{{\vec{\sigma}}_i.{\vec{\sigma}}_{i+1}}+J_2\sum_{i=1}^{N}{{\vec{\sigma}}_i.{\vec{\sigma}}_{i+2}}
\end{equation}	
Here, N represents the number of sites present in the spin chain, $J_1$ and $J_2$ are antiferromagnetic coupling coefficients of nearest and next nearest neighbor interactions and $\vec{\sigma}=\sigma^x\hat{x}+\sigma^y\hat{y}+\sigma^z\hat{z}$ where $\sigma^x, \sigma^y, \sigma^z$ are the Pauli spin matrices. Some solid state systems like $SrCuO_2$ may be described by this Hamiltonian~\cite{MATSUDA19951671}. Periodic boundary condition, $\sigma_{N+1}=\sigma_1$, has been imposed on all systems that have been investigated in this paper. It was known previously that this spin system goes from spin-fluid phase to dimer phase around $\alpha=J_2/J_1\approx0.241$. In the weakly frustrated region, $0<\alpha<0.24$ the system is gapless while it enters a gapped region for higher values of the coupling parameter~\cite{majumdar1969next,PhysRevA.70.052302,PhysRevB.54.9862}.

\begin{figure}
    \includegraphics[width=8.5cm]{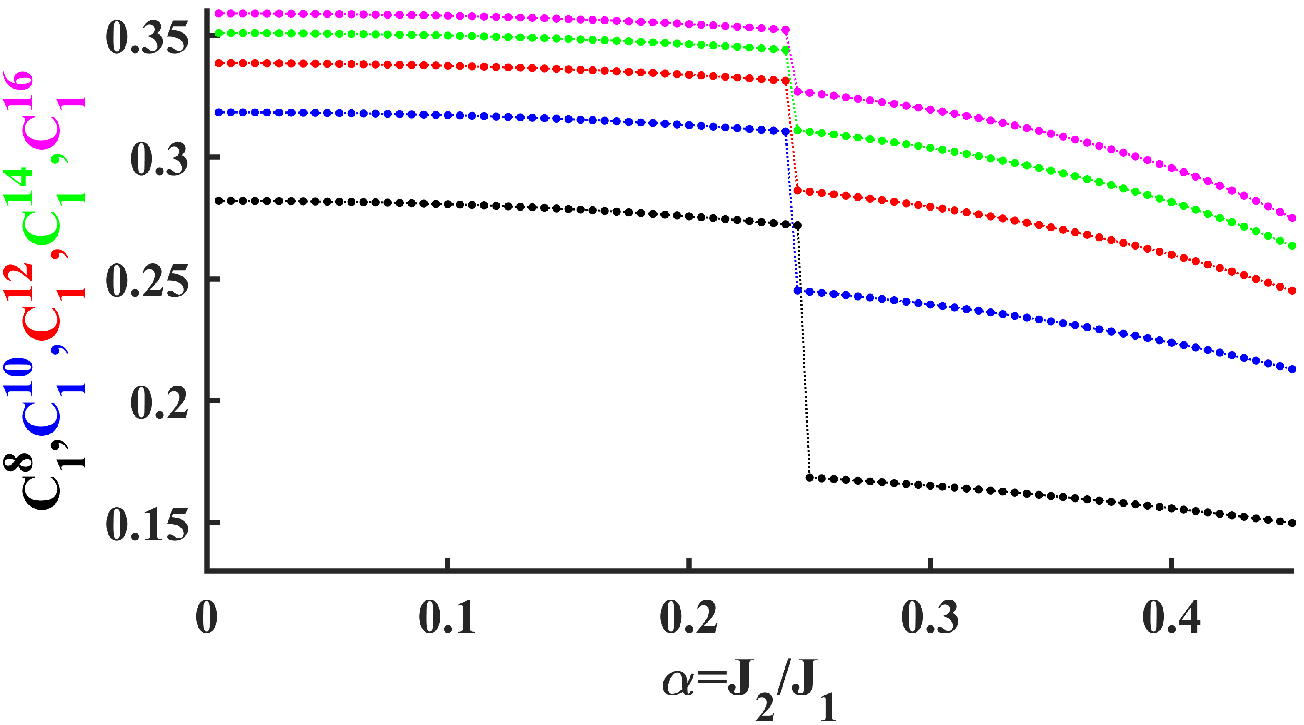}
(a) 
    \includegraphics[width=8.5cm]{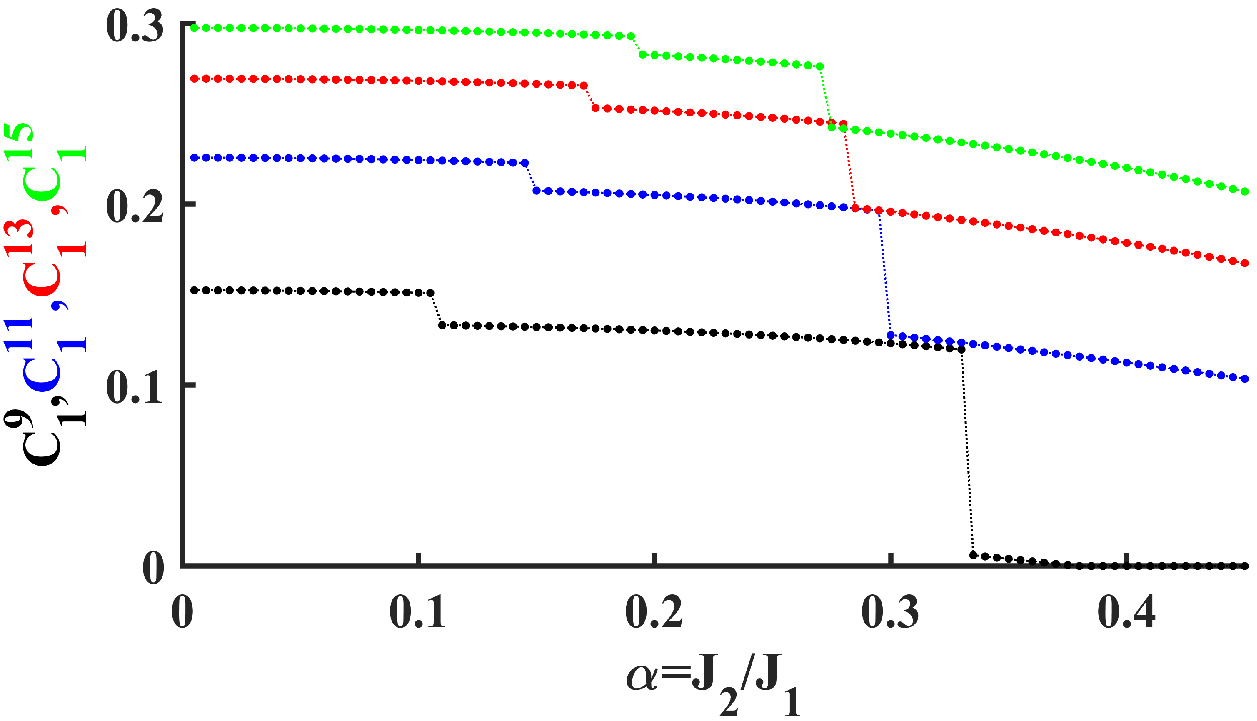}
(b)
  \caption{(color online) Nearest neighbor concurrence in ebits of the first excited state of the $1$-D $J_1-J_2$ Hamiltonian is plotted with respect to the dimensionless system parameter $\alpha$, (a) for spin chains with even number of qubits and (b) for spin chains with odd number of qubits.}
\label{fig:conc1d}
\end{figure}

\begin{figure}[h!]
\includegraphics[width=8.5cm]{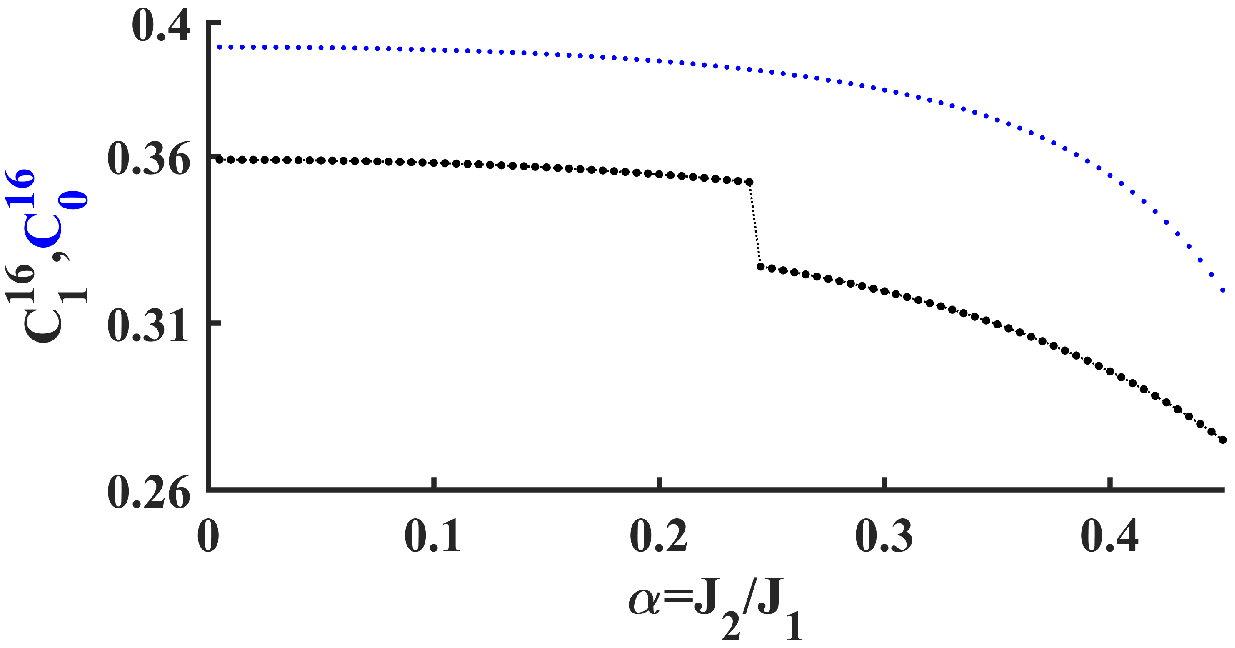}
\caption{(color online) Nearest neighbor concurrence in ebits of the ground state and first excited state of $1$-D $J_1-J_2$ Hamiltonian is plotted with respect to dimensionless system parameter $\alpha$ for system size $N=16$ (using partial ARPACK diagonalization). The solid black dots represent first excited state concurrence $(C_1^{16})$ and smaller blue dots represent ground state concurrence $(C_0^{16})$.} 
\label{fig:conc1d_comp}
\end{figure}

It may be noted that for a two qubit state $\rho$, concurrence $C$ is defined as~\cite{PhysRevLett.78.5022} $\ max(0,\lambda_1-\lambda_2-\lambda_3-\lambda_4)$, where $\lambda_i$’s are the square roots of eigenvalues of $\rho(\sigma^y\otimes\sigma^y)\rho^\ast(\sigma^y\otimes\sigma^y)$ in decreasing order and $\rho^\ast$ is complex conjugate of $\rho$. We perform exact diagonalization of the system Hamiltonian for system sizes $N=8$ to $N=15$. For large spin chains $(N>15)$ we are unable to use the exact diagonalization technique to calculate the eigenvalues and eigenvectors due to memory constraint of the computers used for computation. For $N=16$, we use ARPACK (available in MATLAB that uses lanczos algorithm) to calculate first $6$ low lying eigen states. The results obtained using the Lanczos algorithm is compared with exact diagonalization results for system sizes up to $N=15$ and both the eigenvalues and eigenvectors are found to be fairly accurate. We find the ground state and the low lying excited states and calculate the nearest neighbor concurrence, after tracing out the other qubits. In Fig.~\ref{fig:conc1d}~(a), we plot the nearest neighbor concurrence for the first excited states for the systems with even number of qubits from $N=8$ to $N=16$ and notice discontinuities in the plots in the vicinity of the quantum phase transition point. In Fig.~\ref{fig:conc1d}~(b), we plot the nearest neighbor concurrence for the first excited states for the systems with odd number of qubits from $N=9$ to $N=15$ and notice a pair of discontinuities in the plots in the vicinity of the quantum phase transition point. Finite size scaling analysis is done for data obtained for both even and odd spin chains to ascertain the behavior of the systems as $N\rightarrow\infty$. In Fig.~\ref{fig:conc1d_comp}, we plot the nearest neighbor concurrence of the ground state as well as of the first excited state for the spin chain with $N=16$. The plot of nearest neighbor concurrence for qubits is continuous across the QPT point for the ground state of the system whereas we notice a sudden drop in the value of concurrence in the vicinity of the quantum critical point at $\alpha_c^{16}=0.24248$. The discontinuity of the bipartite entanglement of the first excited state of the system indicates the quantum phase transition point whereas a similar probe applied to the ground state of the same system fails to indicate the quantum phase transition.

\begin{table}[h!]
\begin{ruledtabular}
\begin{tabular}{ccccc}
 $N_{(even)}$&$\alpha_c^{N_{(even)}}$&$N_{(odd)}$&$\alpha_{c,R}^{N_{(odd)}}$&$\alpha_{c,L}^{N_{(odd)}}$\\ 
\hline
 8  & 0.24630 &  9 & 0.10855 & 0.33049 \\
 10 & 0.24449 & 11 & 0.14910 & 0.29944\\
 12 & 0.24349 & 13 & 0.17465 & 0.28243\\
 14 & 0.25288 & 15 & 0.19145 & 0.27199\\
 16 & 0.24248 & & & \\
\end{tabular}
\end{ruledtabular}
\caption{The driving parameter corresponding to the discontinuities in the nearest neighbor concurrence is listed against the appropriate number of qubits.}
\label{tab:table1}
\end{table}

In Table~\ref{tab:table1}, we have listed the values of the driving parameters at the discontinuities of the nearest neighbor concurrence of the first excited states for spin chains of $N=8$ to $N=16$ qubits. Similar results for even number of qubits were found earlier using conformal field theory by K. Okamoto and K. Nomura~\cite{OKAMOTO1992433} for the QPT point, and our calculated values match with their results up to the fourth decimal place. The discontinuities associated with even spin chains are closer to the quantum phase transition point. The numerical values of $\alpha_c^N$ decrease with increasing N, for even number of qubits and asymptotically tend towards a fixed value $\alpha_c$. We fit a rational function $F(N)$ with second degree polynomials in numerator and denominator  through the tabulated values associated with even spin chains; 
\begin{equation}
	F\left(N\right)=\frac{p_1N^2+p_2N+p_3}{N^2+q_1N+q_2}	
\end{equation}
where $p_1=0.2412$, $p_2=0.1477$, $p_3=0.4848$, $q_1=0.6151$, and $q_2=0.5081$. 
\begin{figure}
\includegraphics[width=8.5cm]{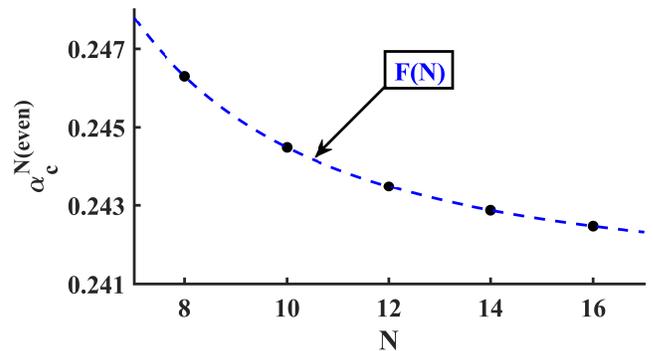}
\caption{(color online) The position of discontinuities of the nearest neighbor concurrence $\alpha_c^{N_{(even)}}$ is plotted with respect to system size N.  $F\left(N\right)$ is the curve fitted through these points.}
\label{fig:extrapolate}
\end{figure}
In Fig.~\ref{fig:extrapolate} we plot the position of discontinuities of the nearest neighbor concurrence $\alpha_c^{N_{(even)}}$ with respect to system size N. We choose the rational function because of its known advantages in extrapolation. The QPT point is estimated to be at $\alpha_c\approx0.24116$ from the extrapolated function.

It may be noted that there are two discontinuities in nearest neighbor concurrence of odd spin chains which appear to asymptotically converge to some point of the system parameter $\alpha$. We have listed the right shifting as well as left shifting discontinuities for spin chains with odd number of particles from  $N=9$ to $N=15$ in Table.~\ref{tab:table1}. To study the convergence of the two discontinuities and to gain further insight into finite size quantum spin systems engineered in the laboratories, we study the scaling of the QPT points with respect to N.

\begin{figure}
    \includegraphics[width=8.5cm]{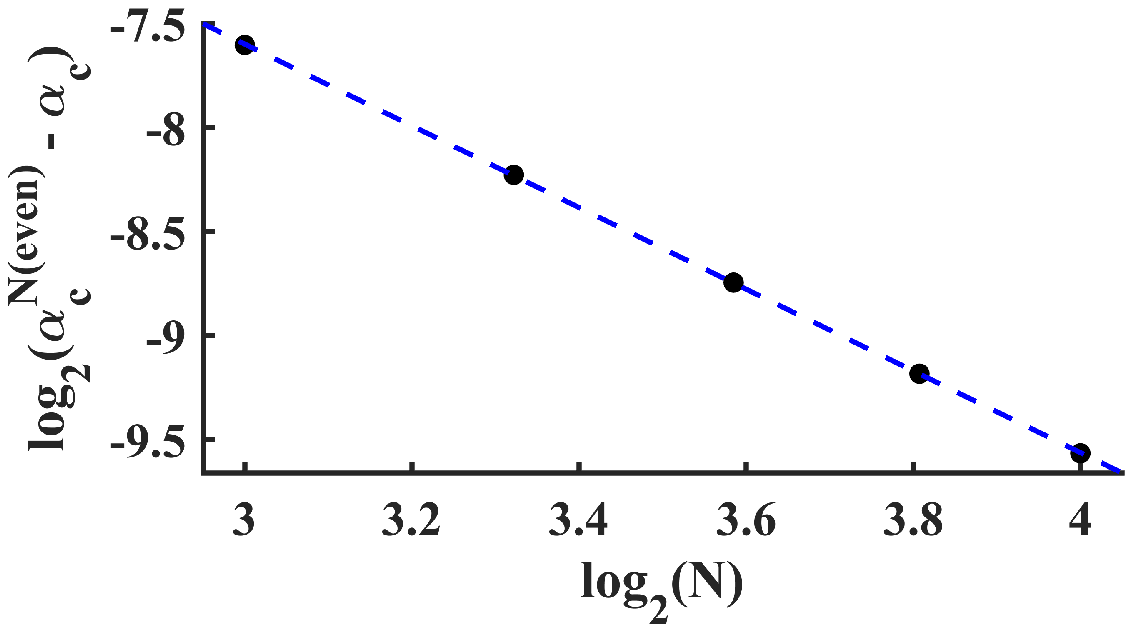}
(a) 
    \includegraphics[width=8.5cm]{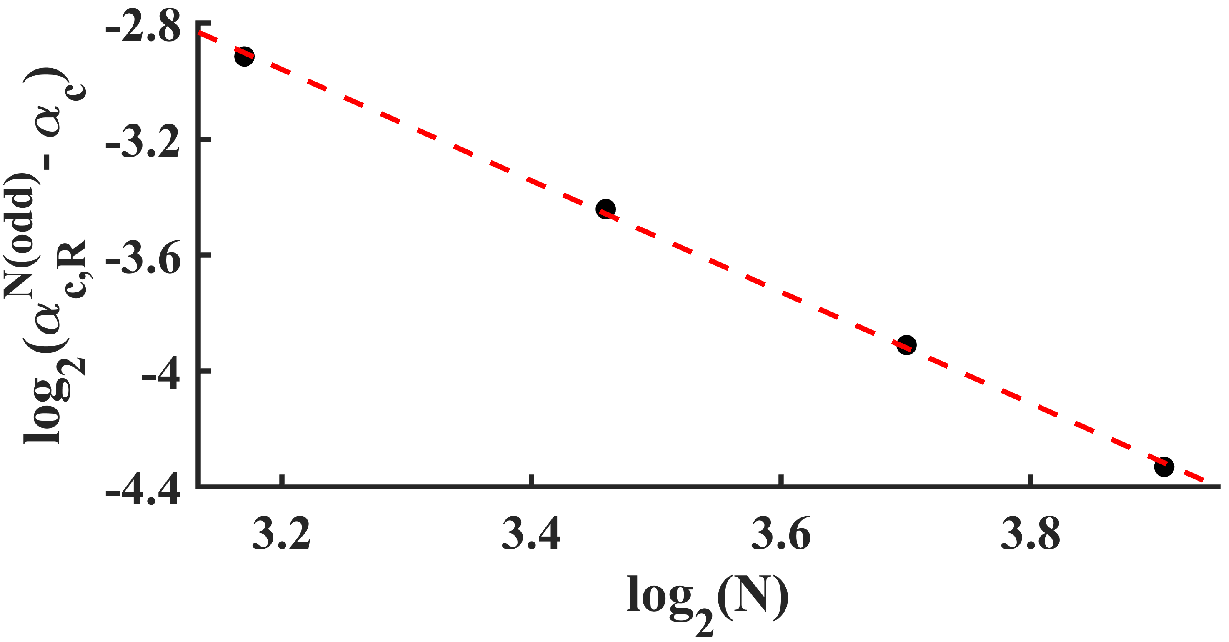}
(b)
    \includegraphics[width=8.5cm]{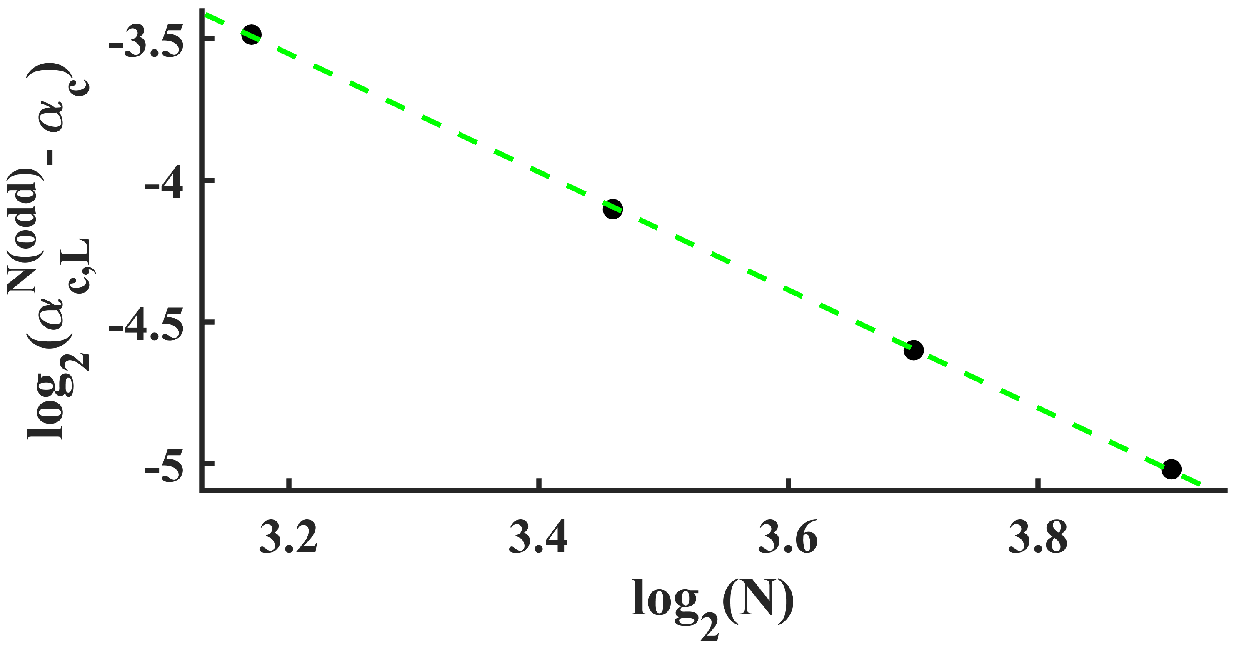}
(c)
  \caption{(color online) (a) $\log_2\left(\alpha_c^{N_{(even)}}-\alpha_c\right)$ versus $\log_2\left(N\right)$ plot, (b) $\log_2\left(\alpha_{c,R}^{N_{(odd)}}-\alpha_c\right)$ versus $\log_2\left(N\right)$ plot and (c) $\log_2\left(\alpha_{c,L}^{N_{(odd)}}-\alpha_c\right)$ versus $\log_2\left(N\right)$ plot.}
  \label{fig:scaling}
\end{figure}

In Fig.~\ref{fig:scaling} (a) we plot $\log_2\left(\alpha_c^{N_{(even)}}-\alpha_c\right)$ with respect to $\log_2\left(N\right)$. We find that a straight line fits the plot, the equation of which is obtained by the method of least squares. The sum of squares due to errors $(SSE)$, which measures the total deviation of the fit from the response values, associated with the straight line fit is $5.5561\times{10}^{-5}$. The equation of the straight line is given by 
\begin{equation}
	\log_2\left(\alpha_c^{N_{(even)}}-\alpha_c\right)\ =\beta{\log_2\left(N\right)}+c	
	\label{eqn:scaling-even}
\end{equation}	
with, $\beta=-1.962$ and $c=-1.715$. 
We may write equation~\ref{eqn:scaling-even}  as 
\begin{equation}
	\alpha_c^{N_{(even)}}=\ \alpha_c+{0.3046\ N}^{-1.962}
	\label{eqn:scaling-even-redef}
\end{equation}	
From the previous equation we note that $\alpha_c^{N_{(even)}}$ approaches $\alpha_c$ as $N^{-1.962}$. The scaling exponent obtained, using this method of detection of QPT point, $\beta=-1.962$ is significantly high. We use the value of $\alpha_c$ obtained by analysing the even spin chains for the scaling analysis of odd spin chains. In Fig.~\ref{fig:scaling} (b) and Fig.~\ref{fig:scaling} (c) we plot $\log_2\left(\alpha_{c,R}^{N_{(odd)}}-\alpha_c\right)$ and $\log_2\left(\alpha_{c,L}^{N_{(odd)}}-\alpha_c\right)$ with respect to $\log_2\left(N\right)$, to study the right-shifting and left-shifting discontinuities of odd spin chains. The SSE associated with the plots are $7.2094\times{10}^{-4}$ and $1.3726\times{10}^{-4}$ respectively and the corresponding equations may be written as
\begin{equation}
	\alpha_{c,R}^{N_{(odd)}}=\ \alpha_c+{9.082\ N}^{-1.92}	           
\end{equation}
\begin{equation}
	\alpha_{c,L}^{N_{(odd)}}=\ \alpha_c+{8.64\ N}^{-2.082}.	           
\end{equation}
The data points for the odd spin chains fit very well in the finite size scaling plot and the right and left shifting discontinuities approach $\alpha_c$ as $N^{-1.92}$ and $N^{-2.082}$ respectively.\\
\begin{figure}
 \includegraphics[width=8.5cm]{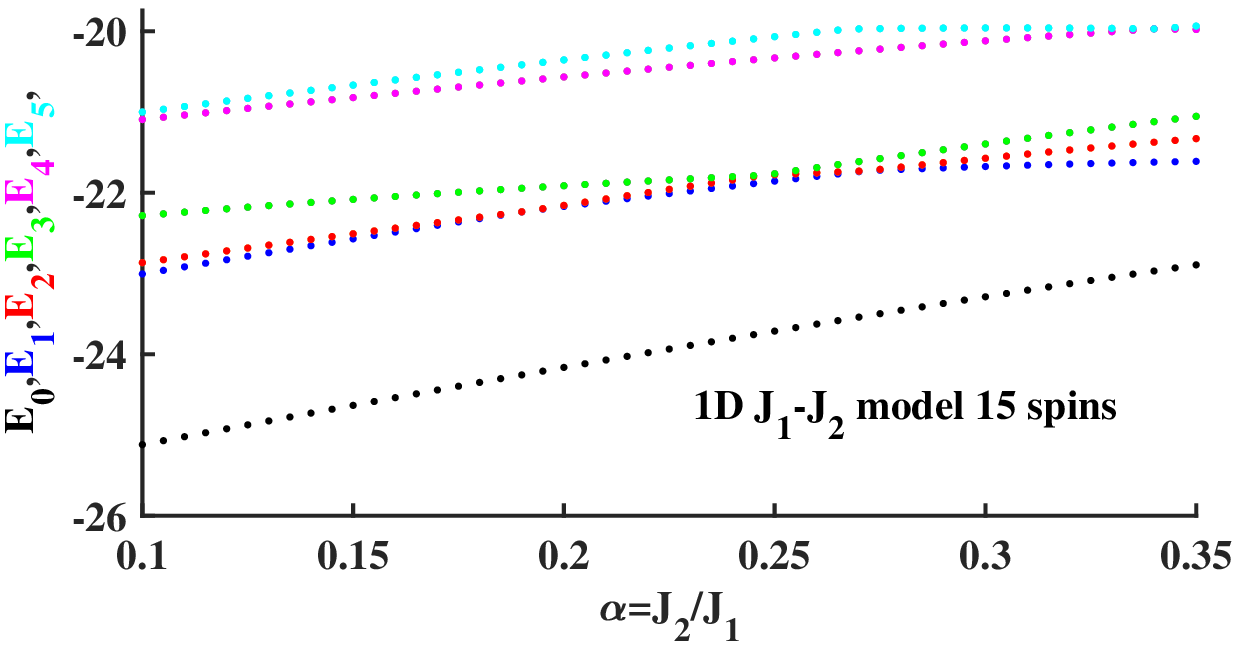}
(a)
 \includegraphics[width=8.5cm]{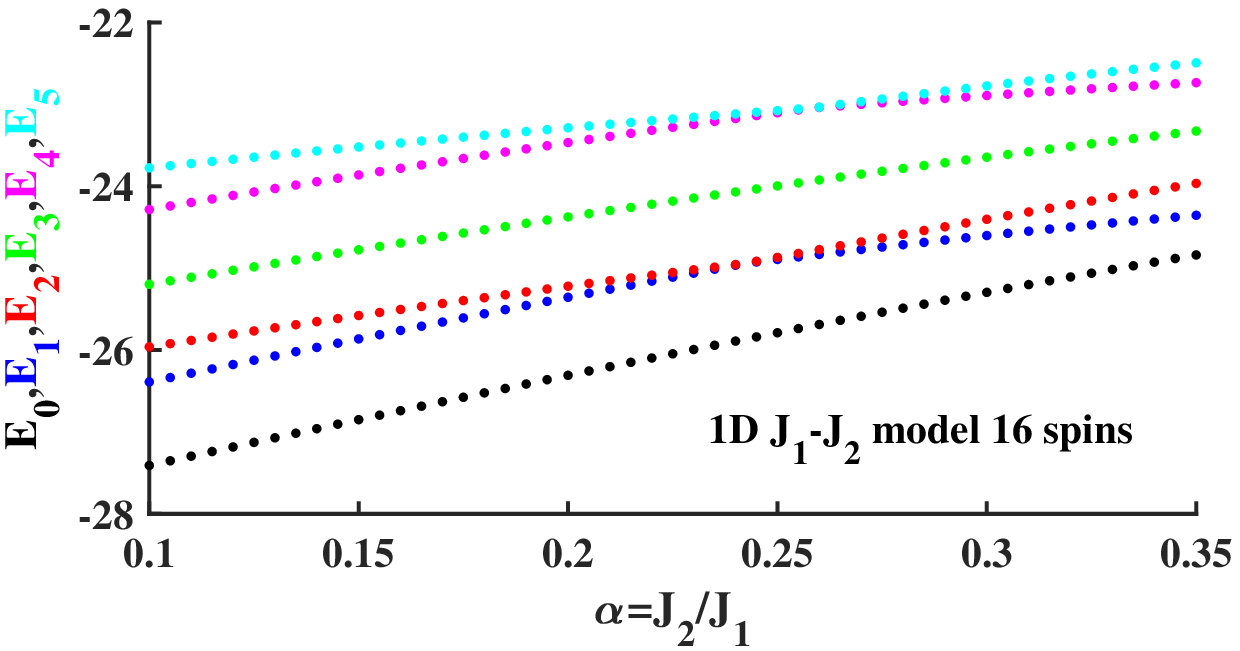}
(b)
 \caption{(color online) Energy levels of ground and low lying excited states are plotted with respect to the dimensionless driving parameter $\alpha$ of the (a) frustrated $J_1-J_2$ spin chain with odd number of qubits, (b) frustrated $J_1-J_2$ spin chain with even number of qubits.}
 \label{eld-15-16}
\end{figure}
It may be noted that a previous investigation of the same spin chain for even number of qubits by Chen~$\etal$~\cite{chen2007} focused on the fidelity of adjacent pure ground states and excited states as a function of the driving parameter $\alpha$. It was reported that the adjacent ground state fidelity practically remains constant and equal to 1 for $0<\alpha<0.5$. However, the fidelity of adjacent first excited states deviates from 1 close to the quantum phase transition point. From the energy level diagram Fig.~\ref{eld-15-16}~(b) it can seen that there is no level crossing for ground state for finite size systems in the parameter range $0\le\alpha\le0.35$. Note that the first and second excited energy levels cross in the vicinity of the quantum phase transition point. In a paper by G.S.Tian and H.Q.Lin~\cite{tian2003}, it was shown that level crossing of low lying excited states causes quantum phase transition when there is no ground state level crossing. For the case of spin chains with odd number of qubits we find two discontinuities and observe that both the discontinuities approach towards the same quantum phase transition point. We may clearly see the reason of the two discontinuities from the energy level diagram Fig.~\ref{eld-15-16}~(a) of 15 qubits spin chain -- there are actually two level crossings between the first and second excited states. In Ref.~\cite{doi:10.1143/JPSJ.56.2153}, the authors had considered the singlet-triplet energy gap as an indicator of quantum phase transition. The phase transition point was determined ($\approx0.3$) by extrapolating the singlet-triplet energy gap for infinite spin chain. The singlet state with energy $E_0^0(N,\alpha)$ is the ground state of the system while the triplet state with energy $E_1^0(N,\alpha)$ is the first excited state of the spin chain before the quantum phase transition point. The triplet state with energy $E_1^0(N,\alpha)$ becomes the second excited state for $\alpha>\alpha_c$. In Ref.~\cite{OKAMOTO1992433} the authors considered the difference between the singlet-singlet energy gap and the singlet-triplet energy gap as an indicator of quantum phase transition. The analysis was done for finite size spin chains. The singlet-singlet energy gap defined in Eq.~\ref{sseqn}, is the difference of energy of the ground and second excited states of the spin system before the quantum phase transition point while it is the difference between the ground and first excited states after the phase transition point. The phase transition point was determined at $\alpha_c=0.2411$ by extrapolation. The method for detection of quantum phase transition, using the entanglement of first excited state as an indicator, relies on the crossing of the first and second excited states for finite size spin chains. The difference between the singlet-singlet energy gap and the singlet-triplet energy gap becomes zero~\cite{OKAMOTO1992433} at the point of intersection of the first and second excited energy levels. The fidelity of the first excited pure state dips at the point of phase transition also due to the intersection of the first and second excited energy levels.

\section{the two DIMENSIONAL HEISENBERG $J_1-J_2$ spin system}
\label{sec3}
We consider an arrangement of qubits in two dimensional square lattice, where the nearest neighbor spins are coupled by Heisenberg interactions, with coupling strength $J_1$ and the next nearest neighbor or diagonal spins are coupled by the same interactions with coupling strength $J_2$. The coupling strengths $J_1$ and $J_2$ are positive. Magnetic materials such as $Li_2VOSiO_4$ and $Li_2VOGeO_4$ can be described by this Hamiltonian~\cite{PhysRevLett.85.1318,PhysRevLett.88.186405,PhysRevB.78.064422,PhysRevLett.101.057010,PhysRevLett.101.076401}. We measure first excited state nearest neighbor concurrence  in a square lattice with $(4\times4)$ sites. The system Hamiltonian is given by
\begin{equation}
	H_{2D}=J_1\sum{{\vec{\sigma}}_i.{\vec{\sigma}}_j+J_2\sum{{\vec{\sigma}}_i.{\vec{\sigma}}_k}}	
\end{equation}
where i, j are nearest neighbors (horizontal or vertical) and i, k are next nearest neighbors or diagonal spins. $J_1$ and $J_2$ are antiferromagnetic. Periodic boundary condition is imposed during computation. The spin model has been studied using exact diagonalization, field theory methods~\cite{richter2010spin,PhysRevLett.80.2705,PhysRevB.79.094413}, but the exact phase boundaries are not known. It is predicted that there are two long range ordered phases separated by quantum paramagnetic phase without long range order, in the system. It has also been predicted that quantum phase transitions exist from ‘ordinary-$N\acute{e}el$ order’ to intermediate phase and from that intermediate phase to ‘colinear-$N\acute{e}el$ order’ at $\alpha\approx0.4$ and $\alpha\approx0.6$ respectively~\cite{Schulz_1992,PhysRevB.51.6151}. The intermediate phase~\cite{PhysRevB.89.241104} is  predicted as plaquette or columnar dimer phase~\cite{PhysRevB.42.8206,PhysRevLett.93.127202,PhysRevB.81.144410,PhysRevB.79.024409,PhysRevB.44.12050,PhysRevB.74.144422,PhysRevB.73.184420,PhysRevLett.91.197202,PhysRevB.78.214415} as well as spin fluid phase~\cite{PhysRevB.86.024424,PhysRevB.88.060402,PhysRevB.86.075111}.
\begin{figure}
\includegraphics[width=8.5cm]{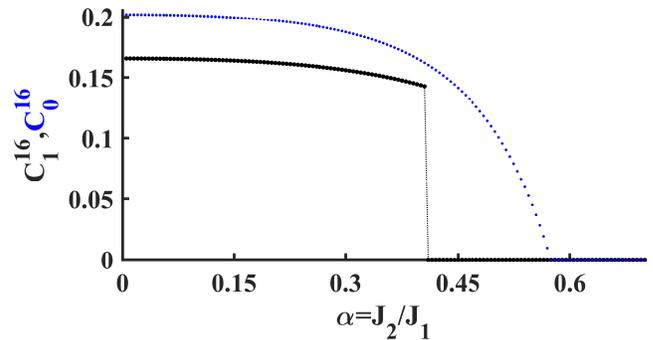}
\caption{(color online) Nearest neighbor concurrence in ebits of the ground state and first excited state of the $2$-D $J_1-J_2$ Hamiltonian is plotted with respect to dimensionless system parameter $\alpha$ for system size $N=16$ (using partial ARPACK diagonalization). The solid black dots represent first excited state concurrence $(C_1^{16})$ and smaller blue dots represent ground state concurrence $(C_0^{16})$.}
\label{fig:2d}
\end{figure}
\begin{figure}
\includegraphics[width=8.5cm]{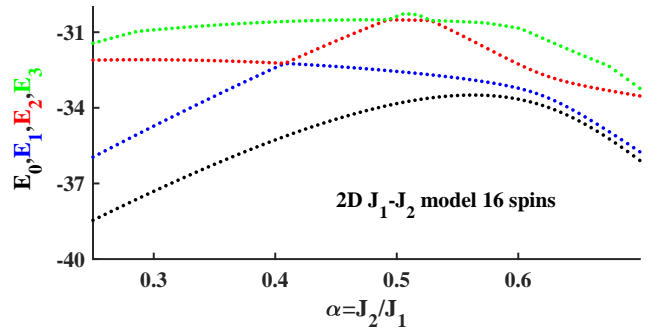}
\caption{(color online) Energy levels of ground and low lying excited states are plotted with respect to the dimensionless driving parameter $\alpha$ of the 2-dimensional frustrated $J_1-J_2$ spin system with 16 $(4\times4)$ qubits.}
\label{fig:eld_2d}
\end{figure}

It can be seen from Fig.~\ref{fig:2d} that the nearest neighbor concurrence of the ground state goes to zero and indicates the intermediate to collinear-$N\acute{e}el$ phase QPT at $\alpha=0.58$. Further, there is a sudden disappearance of nearest neighbor concurrence of the first excited state at $\alpha=0.4078$, indicating the ordinary-$N\acute{e}el$ to intermediate phase QPT point. Note that the ground state nearest neighbor concurrence does not detect the ordinary-$N\acute{e}el$ to intermediate phase QPT point. Similar to the one dimensional case here also we see from the energy level diagram, Fig.~\ref{fig:eld_2d}, that there is a level crossing between the first and second excited states in the vicinity of the quantum phase transition point.

\section{the SHASTRY-SUTHERLAND spin system}
\label{sec4}
\begin{figure}
\includegraphics[height=5cm,width=6cm]{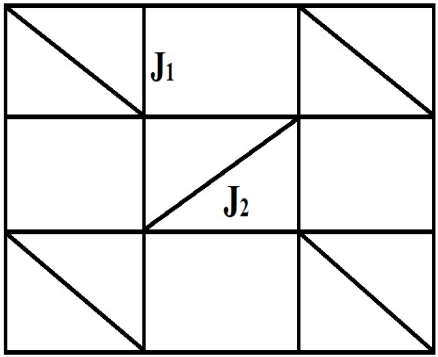}
\caption{The Shastry-Sutherland lattice with $16$ sites. The horizontal and vertical lines represent nearest neighbor coupling strength $J_1$ and the specific diagonal lines represent next nearest neighbor coupling strength $J_2$.}
\label{fig:ss_model}
\end{figure}

We study the entanglement properties of the first excited state of the Shastry-Sutherland quantum spin Hamiltonian for a $(4\times4)$ square lattice, the schematic diagram of which is shown in Fig.~\ref{fig:ss_model}.

The Hamiltonian of the spin system is given by
\begin{equation}
	H_{SS}=J_1\sum{{\vec{\sigma}}_i.{\vec{\sigma}}_j+J_2\sum{{\vec{\sigma}}_k.{\vec{\sigma}}_l}}	
\end{equation}		
where i, j are the nearest neighbors (horizontal and vertical) and k, l are the specific diagonal pairs~\cite{SRIRAMSHASTRY19811069} shown in Fig.~\ref{fig:ss_model}. The coupling strengths $J_1$ and $J_2$ are both positive. Periodic boundary condition is imposed during computation.

It is predicted that the system goes through two quantum phase transitions from $N\acute{e}el$ to intermediate phase and from intermediate phase to dimer, driven by quantum fluctuations~\cite{PhysRevB.64.134407,Albrecht_1996}. The nature of the intermediate phase is not yet known. The quantum phase transition from an intermediate phase to dimer phase has been predicted by bipartite as well as multipartite  entanglement measures applied on the ground state of the system~\cite{PhysRevA.90.032301} at $\alpha\approx1.53$. However, for this system a multipartite entanglement measure, namely the generalised geometric measure applied on the ground state of the system detects both the quantum critical points, from $N\acute{e}el$ to intermediate phase at $\alpha \approx 1.05$ and from intermediate phase to dimer at $\alpha \approx 1.53$.
\begin{figure}
\includegraphics[width=8.5cm]{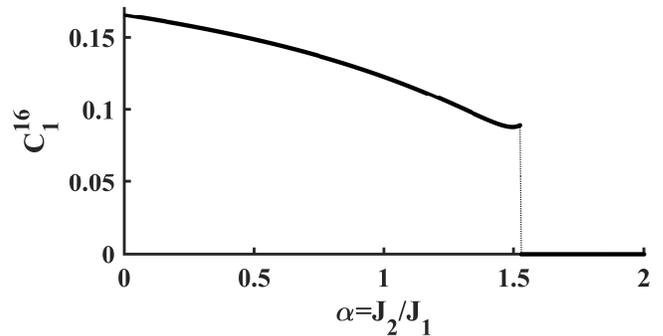}
\caption{(color online) Nearest neighbor concurrence in ebits of the first excited state of the Shastry-Sutherland Hamiltonian  $(C_1^{16})$ is plotted with respect to  the dimensionless driving parameter $\alpha$ for system size $N=16$ (using partial ARPACK diagonalization).} 
\label{fig:ss}
\end{figure}
\begin{figure}
\includegraphics[width=8.5cm]{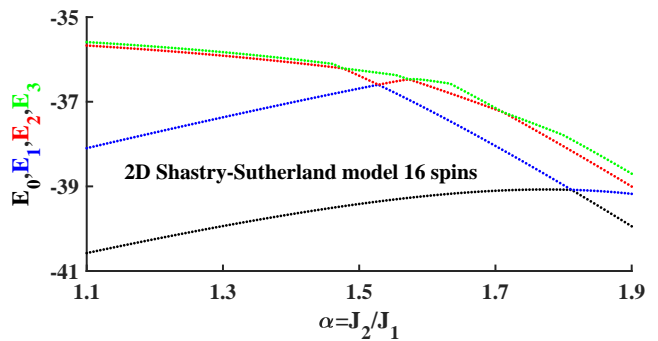}
\caption{(color online) Energy levels of ground and low lying excited states are plotted with respect to the dimensionless driving parameter $\alpha$ of the Shastry-Sutherland spin model with 16 $(4\times4)$ qubits.}
\label{fig:eld_ss}
\end{figure}

In Fig.~\ref{fig:ss}, we note that for $\alpha\geq1.52798$ the nearest neighbor concurrence of the first excited eigenstate of the Hamiltonian $H_{SS}$ $(C_1^{16})$  suddenly becomes zero. The drop in the value of concurrence is sudden indicating the quantum phase transition. Similar to the other models, we see from Fig.~\ref{fig:eld_ss} that there is a level crossing between first and second excited energy levels in the vicinity of quantum phase transition point.\\

\section{CONCLUSIONS}
\label{sec5}
We have investigated the 1D Heisenberg $J_1-J_2$ spin chain, the 2D Heisenberg $J_1-J_2$ spin system and the Shashtry-Sutherland spin system from the viewpoint of bipartite entanglement of their low-lying eigen states. The quantum phase transition points and the phase diagrams of the above mentioned many-body spin systems have often been studied in the past~\cite{PhysRevB.89.241104,OKAMOTO1992433,PhysRevB.54.R9612,PhysRevA.90.032301,PhysRevB.64.134407,Albrecht_1996,PhysRevB.65.014408,PhysRevLett.84.4461}. However, there remains a few unanswered questions regarding the behavior of such systems with respect to their quantum phase diagrams. We find that the bipartite quantum entanglement measure, concurrence of nearest neighbors in first excited states, is discontinuous with the variation of the driving parameter across the quantum phase transition points. It is pertinent to mention that no physical system may be cooled to absolute zero and the low lying excited states of the system are likely to influence the behavior of the system near absolute zero temperature. Furthermore, entanglement entropy has been measured in quantum many-body systems~\cite{RIslam2015} and detection of various quantum entanglement measures have also been reported~\cite{MLi2013}, making the present study viable for experimental investigation.\\
It may be noted that bipartite and multipartite quantum entanglement measures applied on the ground state of the system Hamiltonian~\cite{PhysRevA.90.032301} are unable to detect the quantum phase transition point for the 1D Heisenberg $J_1-J_2$ spin system. The  finite size scaling exponents, obtained using the present investigation for the 1D Heisenberg $J_1-J_2$ model, are also quite high. The investigation of low-lying excited states of such many-body Hamiltonians promises to shed more light on the behavior of quantum spin systems.
\begin{acknowledgments}
We sincerely acknowledge the ``PARAM-Kanchenjunga High Performance Computer Centre, National Institute of Technology Sikkim'' for the support to conduct this research. We thank Ujjwal Sen for critical comments.
\end{acknowledgments}
\bibliography{my_references}
\end{document}